\RequirePackage{amsthm}

\documentclass[sn-chicago]{sn-jnl}

\usepackage{graphicx}%
\usepackage{multirow}%
\usepackage{amsmath,amssymb,amsfonts}%
\usepackage{amsthm}%
\usepackage{mathrsfs}%
\usepackage[title]{appendix}%
\usepackage{xcolor}%
\usepackage{textcomp}%
\usepackage{manyfoot}%
\usepackage{booktabs}%
\usepackage{algorithm}%
\usepackage{algorithmicx}%
\usepackage{algpseudocode}%
\usepackage{listings}%
\usepackage{soul}
\usepackage{subcaption}
\usepackage{tikz}
\newcommand\independent{\protect\mathpalette{\protect\independenT}{\perp}}
\def\independenT#1#2{\mathrel{\rlap{$#1#2$}\mkern2mu{#1#2}}}
\usepackage{bm}
\newcommand{\xx}{1}

\theoremstyle{thmstyleone}%

\theoremstyle{thmstyletwo}%
\newtheorem{example}{Example}%

\theoremstyle{thmstylethree}%

\begin{document}

\title[Staged trees for discrete longitudinal data]{Staged trees for discrete longitudinal data}


\author*[1]{\fnm{Jack Storror} \sur{Carter}}\email{jack.carter@dima.unige.it}

\author[2]{\fnm{Manuele} \sur{Leonelli}}\email{manuele.leonelli@ie.edu}

\author[1]{\fnm{Eva} \sur{Riccomagno}}\email{riccomagno@dima.unige.it}

\author[3]{\fnm{Alessandro} \sur{Ugolini}}\email{alessandro.ugolini@unige.it}

\affil[1]{\orgdiv{Dipartimento di Matematica, Universit\`a degli Studi di Genova, Genova, Italy}}

\affil[2]{\orgdiv{School of Science and Technology, IE University, Madrid, Spain}}

\affil[3]{\orgdiv{Department of Surgical and Diagnostic Sciences, University of Genoa, Genoa, Italy}}

\abstract{In this paper we investigate the use of staged tree models for discrete longitudinal data.  Staged trees are a type of probabilistic graphical model for finite sample space processes.  They are a natural fit for longitudinal data because a temporal ordering is often implicitly assumed and standard methods can be used for model selection and probability estimation.  However, model selection methods perform poorly when the sample size is small relative to the size of the graph and model interpretation is tricky with larger graphs.  This is exacerbated by longitudinal data which is characterised by repeated observations.  To address these issues we propose two approaches: the longitudinal staged tree with Markov assumptions which makes some initial conditional independence assumptions represented by a directed acyclic graph and marginal longitudinal staged trees which model certain margins of the data.}




\maketitle

\section{Introduction}

Longitudinal studies are characterised by the repeated observation of individuals over time.  This differs to cross-sectional studies in which each individual is only observed at a single time point.  Longitudinal studies benefit from increased power over cross-sectional studies \citep{zeger1992overview} and are able to answer more sophisticated questions because the data shows how individuals change over time.  However, they require additional modeling considerations due to the dependence between the repeated observations.

Longitudinal studies are often viewed in a regression context in which one variable is considered the outcome and the remaining variables the covariates.  The aim is then to determine the effect of the covariates on the outcome.  We denote the repeated observations of the \textit{longitudinal outcome} by $Y_t$, $t=1\dots,T$ and of the \textit{longitudinal covariates} by $X_t$, $t=1\dots,T$ where $T$ is the total number of time points.  The study may also include covariates whose value does not change over time.  We call these \textit{time invariant covariates} and denote them by $Z$.  In this paper we assume that all variables are discrete with a finite sample space.  The finite sample space may be either categorical or ordinal.

Traditional methods for such longitudinal data typically involve a generalised linear model with three approaches being particularly popular - the marginal model, transition model and random effects model.  Each of these approaches offer different methods for modeling both the effect of the covariates on the outcome and the dependence between the outcome at different time points.  Further details of these models will be given in Section \ref{sec:Comparison}.

In this paper, we instead investigate the use of staged tree models for discrete longitudinal data.  Staged trees, first introduced by \cite{smith2008conditional} (see \cite{collazo2018chain} for a detailed review) are a type of probabilistic graphical model represented by a coloured directed tree graph.  They also have a more compact graphical representation called the chain event graph, obtained through a coalescence of the vertices.  The main feature of staged tree models is the ability to identify events that have identical probability distributions.  For example, in a regression setting they can specify different values of the covariates for which the probability distribution of the outcome is the same.  Additionally, a temporal ordering of variables is often implicitly assumed in staged tree models and so longitudinal data easily fits into the staged tree framework.

While including a temporal element has been explored in staged trees - for example, the dynamic chain event graph and the inclusion of conditional holding times \citep{barclay2015dynamic} - and longitudinal data has been modelled using staged trees in applications \citep{hutton5chain}, no specific methods for using staged trees with the form of longitudinal data described above have been proposed.  Staged trees are often compared to Bayesian networks - another type of probabilistic graphical model - for which some methods for longitudinal data have been proposed.  For example, the dynamic Bayesian network for longitudinal data \citep{McGeachie2015Longitudinal,Prinzie2011Modeling,Chen2012Dynamic} and improved probability estimation for longitudinal data \citep{Bellazzi1998Learning}.

Regression based methods make efficient use of the data by making strong assumptions - namely that the effect of the covariates on the outcome follows a specific linear model and the dependence between outcomes at different times follows a specific form.  The inferences obtained often rely on these assumptions which can at times be hard to interpret or test.  When selected by data, staged tree models do not require any such initial assumptions.  However, this comes with the drawback that they require a large sample size relative to the number of variables and the size of their sample spaces.  This is critical with longitudinal data because observing an additional time point results in an increase in the number of variables in the model without an increase in the sample size.  

While existing staged tree methodology can be applied to longitudinal data, in this paper we suggest two additional approaches for smaller sample sizes.  One makes some initial conditional independence assumptions which can be represented via a directed acyclic graph.  These assumptions are easy to interpret due to their graphical representation and could also be tested using data via standard methods for Bayesian networks \citep[see e.g.][]{scutari2021bayesian}.  The other approach models certain marginal distributions with staged trees.  While this no longer allows estimation of the full joint probability distribution, one is still able to determine how the effect of the covariates on the outcome changes over time.

The remainder of the paper is organised as follows.  In Section \ref{sec:STs} we introduce staged tree models and give a brief summary of current model selection methods for staged trees.  Section \ref{sec:LongSTs} introduces three proposed approaches for using staged trees to model longitudinal data - the full longitudinal staged tree, the longitudinal staged tree with Markov assumptions and marginal longitudinal staged trees.  In Section \ref{sec:Comparison} we compare staged tree models with traditional regression methods and discuss some of the benefits of using staged tree models for longitudinal data.  We finish in Section \ref{sec:Caries} by applying the proposed staged tree methods to a real data set about tooth decay in children.

\section{Staged trees and chain event graphs}\label{sec:STs}

\subsection{Definition}

In this section we present staged tree models and their compact representation chain event graphs with the help of an example.  This example will later be extended to demonstrate our proposed methods for longitudinal data.  The example first appeared in \cite{koch1977general} and it compares a new drug for treating depression with a standard treatment.  Individuals were randomly assigned one of the treatments and were also diagnosed to have either mild or severe depression.  After one week of treatment the individual's symptoms were assessed to be either normal (N) or abnormal (A).

\begin{figure}
    \centering
    \begin{tabular}[c]{c} 
        \subcaptionbox{Event tree with observed counts\label{subfig:et}}
        {\includegraphics[scale=0.5]{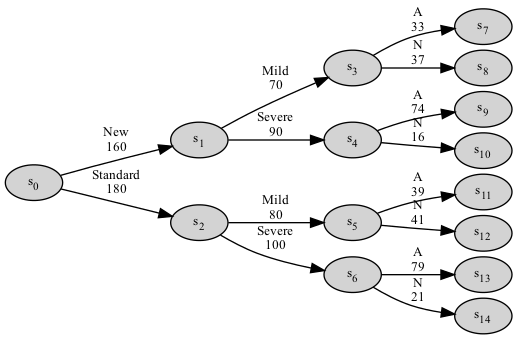}} \\ \\
        \subcaptionbox{Staged tree with observed counts\label{subfig:st}}
        {\includegraphics[scale=0.5]{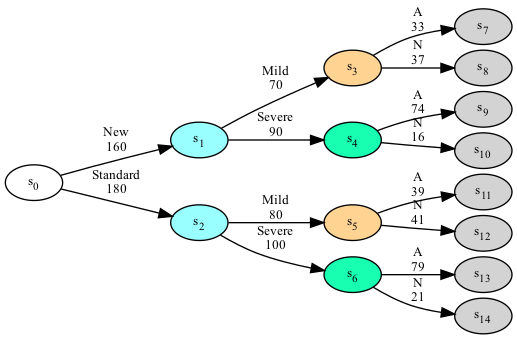}} \\ \\
        \subcaptionbox{Chain event graph with estimated probabilities\label{subfig:ceg}}
        {\includegraphics[scale=0.5]{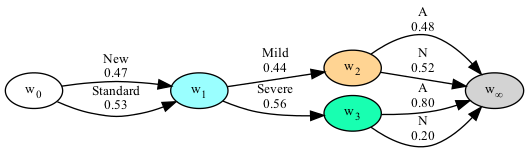}}
    \end{tabular}
    \caption{Construction of a CEG for the single time point depression example.}
\end{figure}

A \textit{staged tree} (ST) is a probability tree with an equivalence relation on its non-leaf vertices.  More specifically, a ST is a probabilistic graphical model for a process consisting of a sequence of discrete events.  To construct a ST we begin with an \textit{event tree} containing a single \textit{root} vertex with no incoming edges and at least two outgoing edges, representing the start of the process, and a number of \textit{leaf} vertices which have a single incoming edge and no outgoing edges, representing the end of the process.  An event tree for the depression data is in Figure~\ref{subfig:et} where the root vertex is labelled $s_0$ and the leaf vertices are $s_7-s_{14}$.  All remaining vertices in the graph have a single incoming edge and at least two outgoing edges.

All non-leaf vertices (including the root) are called \textit{situations} and represent a possible state at which the process can arrive.  The edges in the event tree are labeled such that for each situation, the outgoing edge labels describe all possible events that can occur at the next stage of the process.  As such, the event tree fully describes the sample space of the process.  For example, the vertex $s_2$ in Figure~\ref{subfig:et} represents the situation in which an individual has been assigned to the standard treatment.  The subsequent event is the diagnosis which can be either Mild or Severe.  Edges in Figure~\ref{subfig:et} are additionally labelled with the number of observations of each event.

For each situation in the graph, one can associate a probability distribution representing the conditional probabilities of the subsequent event of the process.  One obtains a joint distribution for the whole process by assigning a probability distribution to all situations and then using the standard chain rule of probability.

The most general statistical model places no further constraints on the probability distributions at each situation.  However, a ST model restricts the space by assuming that some situations (with the same outgoing edge labels) have the same probability distribution.  When this is the case, the two situations are said to be in the same \textit{stage}.  This is represented graphically by colouring vertices according to which stage they are in (with any singleton stages coloured white for simplicity).  For example, in Figure~\ref{subfig:st} the situations $s_1,s_2$ are in the same stage.  This represents that the probability of an individual having mild or severe symptoms does not depend on their treatment assignment - in other words, the treatment assignment and diagnosis are independent.  This would be the case if the treatments have been suitably randomised.  Additionally the situations $s_3,s_5$ are in the same stage, as well as $s_4,s_6$, which represents that the symptoms after one week are independent of the treatment type, conditional on the diagnosis severity.

A \textit{chain event graph} (CEG) is a more concise graphical representation of a ST model.  All leaf vertices are combined into a single \textit{sink} vertex and any two vertices with identical probability distributions for the remainder of the process are combined - such vertices are said to be in the same \textit{position} \citep[for more information on transforming a ST into a CEG see][]{shenvi2020constructing}.  In Figure~\ref{subfig:ceg} the sink vertex is labelled $w_\infty$ and the situations $s_3,s_5$ and $s_4,s_6$ have been merged into the two vertices $w_2$ and $w_3$.  Additionally, the situations $s_1,s_2$ are in the same position so are combined into a single vertex $w_1$.  Notice that the colours of the vertices in the CEG are inherited from the ST.  In Figure~\ref{subfig:ceg}, the edges of the CEG have also been labelled with estimated probabilities.

ST models can be seen as a generalisation of Bayesian networks (BNs).  For example, the ST in Figure~\ref{subfig:st} is equivalent to the BN with a single edge between the diagnosis severity and the symptoms.  In fact, the class of ST models contains all possible BNs \citep{smith2008conditional}.  However, STs are a much richer class of models for two reasons:  
\begin{itemize}
    \item STs are able to specify non-symmetric relationships between variables.  For example, consider the adaptation of the ST in Figure~\ref{subfig:st} in which $s_3,s_5$ are still in the same stage but $s_4,s_6$ are no longer in the same stage.  This represents that the two treatments perform equally for those with mild depression but differently for those with severe depression.  This relationship can be uncovered by a ST but not a BN. Although extensions of BNs representing non-symmetric types of independence have been proposed \citep[e.g.][]{ankinakatte2015modelling,jaeger2006learning,pensar2015labeled}, these often lose the intuitiveness of BNs and no software to learn them from data is freely available.
    \item BNs require that the data be defined via a collection of variables - in STs this is referred to as stratified data.  However, STs can also be used with more complicated forms of data - so called non-stratified data.  Non-stratified data is associated to non-symmetric event trees and can occur due to structural zeros or processes where the set of possible future events depends on past events.  In this paper we will focus on the stratified case but mention briefly the non-stratified case in Section \ref{sec:Comparison}.
\end{itemize}

\subsection{Learning STs from data}

There are two main approaches to using STs with data.  The first assumes a fixed ST, based on expert or prior knowledge, and estimates transition probabilities under this ST.  This helps to improve estimation because samples can be shared across all situations in the same stage which is particularly beneficial when sample sizes are small.  For example, in Figure~\ref{subfig:st}, because $s_1,s_2$ are in the same stage, to estimate probabilities of mild or severe diagnosis we can use all samples from the data set, rather than calculating probabilities separately for those on the new and standard treatments.

The second approach is to perform model selection and learn the staging, and consequently non-symmetric types of independence, from data.  Model selection of STs is often performed using Bayesian methods with a model prior over the space of STs and Dirichlet priors on the transition probabilities.  The model space prior is often chosen to be uniform while a default choice for the Dirichlet priors is based on a score-matching principle and only requires specification of an equivalent sample size \citep[see][for more details]{collazo2018chain}.  The model with highest posterior probability is then selected which can be done by comparing Bayes factors between models.  However, due to the size of the space of ST models, it is often too computationally expensive to perform an exhaustive comparison of all models, hence search algorithms have been proposed to explore the space and approximate the highest posterior model.  The first proposed algorithm, introduced in \citet{freeman2011bayesian}, is called the \textit{agglomerative hierarchical clustering} (AHC) algorithm.  The algorithm starts with a staged tree with each situation in its own stage and sequentially finds the two best stages to be merged, where best means highest Bayes factor. The algorithm continues until no increase in Bayes factor is found.  Throughout this paper we will perform model selection with the AHC algorithm using the \texttt{cegpy} Python library \citep{walley2023cegpy}.

More nuanced structural learning algorithms have since been developed \citep[e.g][]{collazo2016new,leonelli2022structural,silander2013dynamic}. Most often, these search for the staged tree optimizing the model BIC \citep{gorgen2022curved}, using hill-climbing greedy algorithms. The R package \texttt{stagedtrees} \citep{Carli2022} implements a wide array of these routines. 

The two approaches of a fixed ST and model selection can be combined, as will be proposed in Section \ref{sec:Markov}, by selecting the best ST using data-driven routines that respects a set of domain-based assumptions. This is analogous to blacklisting and whitelisting edges in BNs as for instance implemented in the \texttt{bnlearn} R package \citep{Scutari2010learning}.

A key theme of this paper is of a small sample size relative to the size of the event tree.  In this case it is likely that some paths or situations in the event tree will have zero observations.  In such cases, we propose a slight adaptation to the model selection process and the graphical representation of the selected CEG.  First, all zero sample size situations are placed in the same stage (where edge labels allow for this).  Note that these zero sample size stages are fundamentally different to the usual definition of a stage - they do not represent equality of probability distributions but instead the property of having no observed samples in the data set.  Then, when performing model selection, we do not allow these zero sample size stages to be combined with any other stage.  This can be achieved in the \texttt{cegpy} package using a combination of the \texttt{initial\_staging} and \texttt{hyperstage} arguments.  In the selected CEG, we represent the zero sample size stages by a square vertex coloured in grey and any edges with zero sample size with a grey dotted line.

Performing model selection and representing the CEG in this way has two advantages:
\begin{itemize}
    \item Collecting all zero sample size situations together can greatly reduce the size of the resulting CEG and improve readability.
    \item Often the edges of a CEG are labelled with estimated probabilities.  When estimated using Bayesian methods, this will usually be the maximum a posteriori estimate.  However, in the case of zero samples, the posterior distribution is equal to the prior.  By explicitly representing zero sample size situations in the graph, one can easily identify when estimated probabilities are just an artifact of the prior distributions rather than informed by data.
\end{itemize}

\section{Longitudinal ST models}\label{sec:LongSTs}

We now introduce our three proposed methods for modelling longitudinal data with STs.  We will demonstrate the proposed methods on a longitudinal version of the depression example introduced in Section \ref{sec:STs}.

\subsection{Full longitudinal ST}\label{sec:FullST}

Longitudinal data fits naturally into a ST structure.  This is because a temporal ordering of the events in the tree is often implicitly assumed - events that occur later in time appear later in the tree.  Using a ST to model discrete longitudinal data is therefore just a matter of placing all events in a tree with the only modeling choice being the ordering of the variables that are not specified by the temporal ordering.  Most commonly this will place time invariant covariates at the beginning of the event tree \citep{ankinakatte2015modelling} and longitudinal covariates before their associated outcome so that the total ordering is $(Z,X_1,Y_1,\dots,X_T,Y_T)$.  Standard ST methodology can then be used for probability estimation or model selection.  

With longitudinal data, situations associated with the same variable but at different times can be in the same stage.  This can, for example, show when the probability distribution of a variable does not change over time.  Such model selection is possible using the \texttt{cegpy} Python package in which any two situations with the same sample space can be selected to be in the same stage.  However, the \texttt{stagedtrees} R package only allows situations associated to the same variable at the same time to be in the same stage so cannot learn such stagings.

\begin{example}
In the depression example from Section \ref{sec:STs}, symptoms are now assessed to be either normal (N) or abnormal (A) after 1, 2 and 4 weeks of treatment.  The data set is summarised by the event tree in Figure~\ref{fig:DepET} with edges additionally labelled with observed counts.

\begin{figure}
    \centering
    \includegraphics[scale=0.398]{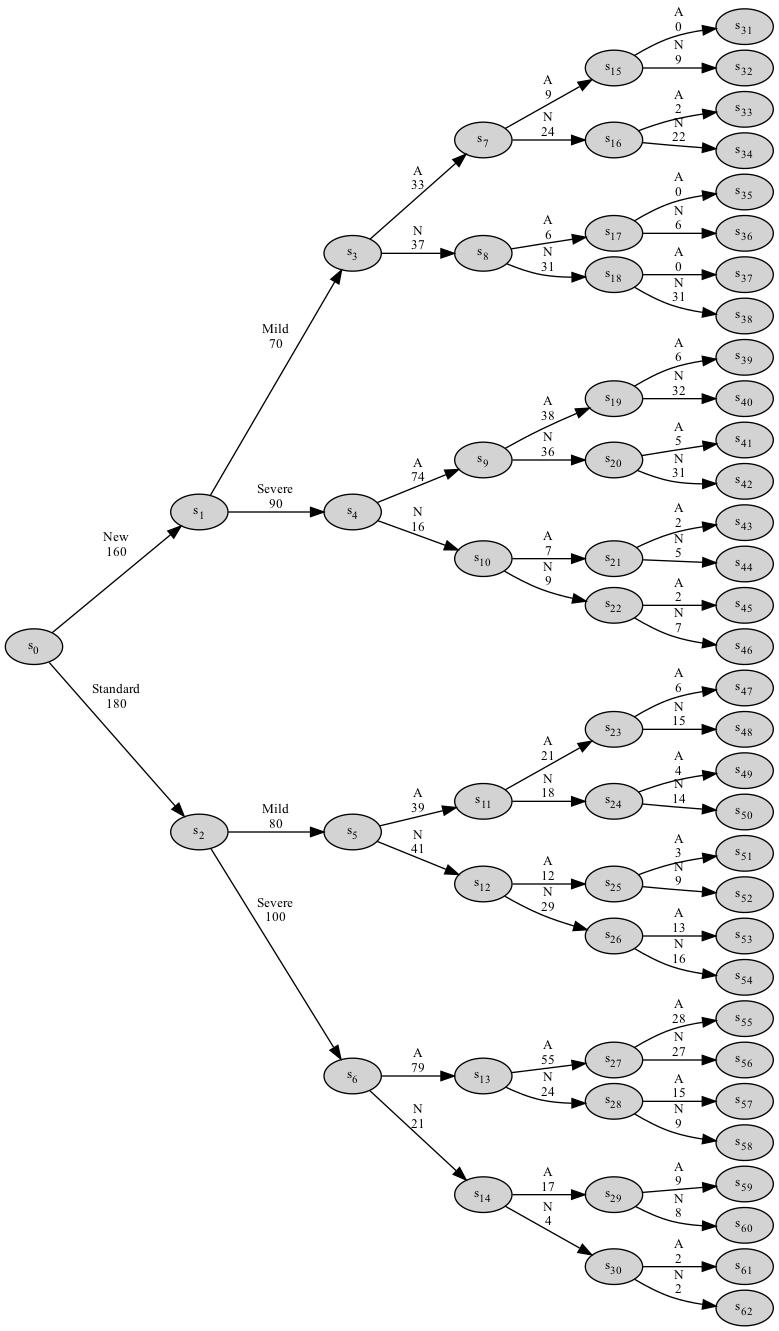}
    \caption{Event tree for depression example.  Edges are labelled with observed counts.}
    \label{fig:DepET}
\end{figure}

The highest posterior probability model is displayed in Figure~\ref{fig:DepCEG}.  Situations related to outcome variables have been grouped into six stages from the pink stage with highest probability of normal symptoms equal to 1 to the green stage with the lowest probability equal to 0.24.  In particular, in week 1 there is no difference between the two treatments - those with mild depression have 0.51 probability of normal symptoms while those with severe depression only have 0.24 probability.  The differences become evident in weeks~2 and 4.  For those with severe depression, taking the standard treatment results in the probability of normal symptoms remaining at 0.24 in week~2 and this only increases to 0.51 in week~4 for some individuals.  However, taking the new drug results in the probability increasing to 0.51 in week~2 and increasing further to either 0.73 or 0.86 in week~4.  Similar can be said for those with mild depression, with the new drug resulting in higher probabilities of normal symptoms in weeks 2 and 4.

\begin{figure}
    \centering
    \includegraphics[scale=0.45]{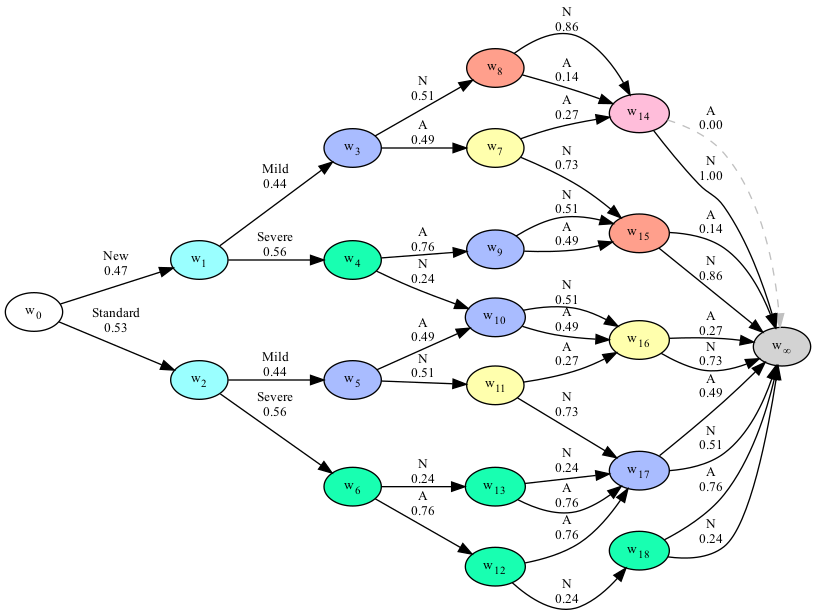}
    \caption{CEG for depression example.  Edges are labelled with estimated probabilities.}
    \label{fig:DepCEG}
\end{figure}

\end{example}

This example shows how STs can be useful for analysing longitudinal data and the sort of conclusions that can be made.  The main advantage of staged tree models is their ability to find context specific properties.  This is even more powerful in longitudinal STs because situations related to the outcome at different times are able to be in the same stage.  For example, in certain situations when using the standard treatment, the probability of normal symptoms is the same in week 1 for those with mild depression as it is in week 4 for those with severe depression.

However, some limitations of this approach are related to the size of the event tree and the sample size.  As the number of paths in the graph increases (due to more time points, additional covariates or outcomes with larger sample space), interpretation of the resulting CEG can be challenging.  Furthermore, the larger the tree, the larger the sample size required to ensure an adequate number of observations at each situation in the tree.  This is already somewhat evident in the above example where the situation $s_{30}$ has only four observations.

\subsection{Longitudinal ST with Markov assumptions}\label{sec:Markov}

One way to utilise a small sample size more efficiently in staged tree models is to assume that certain situations are in the same stage i.e. they are assumed to have the same probabilities of the next event.  Hence, the samples of these situations can be shared in the estimation of their probabilities.  This can also be seen as reducing the number of parameters in the model.  After this initial staging is assumed, additional model selection can be performed. This strategy is adopted by~\citet{leonelli2022highly} to limit the number of model parameters by enforcing an upper bound on the number of allowed dependencies.  This procedure can be performed in the \texttt{cegpy} package using the \texttt{initial\_staging} argument - a feature that has been added to the package for the purposes of this paper.

A simple way to specify which situations to place in the same stage is through conditional independence assumptions of the variables, which can be represented via the Markov assumptions of a directed acyclic graph (DAG).  The choice of Markov assumptions could depend on the specific application and the aims of the analysis and be based on expert knowledge.  Alternatively, one may follow the approach introduced in \citet{barclay2013refining} by first selecting the DAG using data.  A benefit of representing the assumptions by a DAG is that these can be easily converted to a ST using the algorithm of \cite{varando2021staged} implemented in \texttt{stagedtrees}.

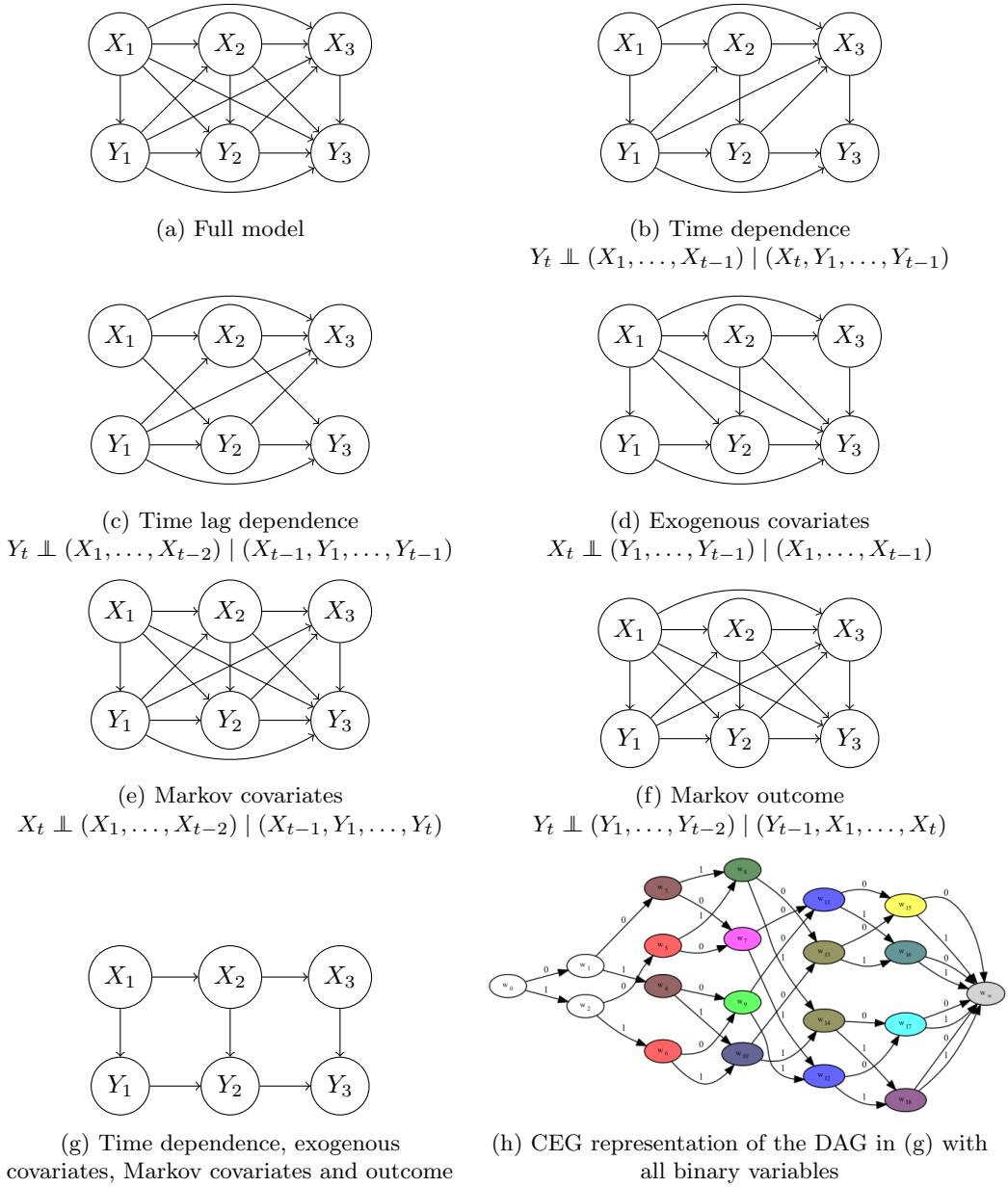
\begin{figure}[]
\begin{tabular}{cc}
\subcaptionbox{\centering Full model\label{subfig:DAG1}}[0.47\linewidth]{
\begin{tikzpicture}[node distance={15mm}, main/.style = {draw, circle}] 
\node[main] (x1) {$X_1$}; 
\node[main] (x2) [right of=x1] {$X_2$};
\node[main] (x3) [right of=x2] {$X_3$};
\node[main] (y1) [below of=x1] {$Y_1$};
\node[main] (y2) [below of=x2] {$Y_2$};
\node[main] (y3) [below of=x3] {$Y_3$};
\draw[->] (x1) -- (y1);
\draw[->] (x1) -- (x2);
\draw[->] (x1) -- (y2);
\draw[->] (x1) to [out=30,in=150] (x3);
\draw[->] (x1) -- (y3);
\draw[->] (y1) -- (x2);
\draw[->] (y1) -- (y2);
\draw[->] (y1) -- (x3);
\draw[->] (y1) to [out=330,in=210] (y3);
\draw[->] (x2) -- (y2);
\draw[->] (x2) -- (x3);
\draw[->] (x2) -- (y3);
\draw[->] (y2) -- (x3);
\draw[->] (y2) -- (y3);
\draw[->] (x3) -- (y3);
\end{tikzpicture}
}
&

\subcaptionbox{\centering Time dependence $Y_t \independent (X_1,\dots,X_{t-1}) \mid (X_t, Y_1, \dots, Y_{t-1})$\label{subfig:DAG2}}[0.53\linewidth]{
\begin{tikzpicture}[node distance={15mm}, main/.style = {draw, circle}] 
\node[main] (x1) {$X_1$}; 
\node[main] (x2) [right of=x1] {$X_2$};
\node[main] (x3) [right of=x2] {$X_3$};
\node[main] (y1) [below of=x1] {$Y_1$};
\node[main] (y2) [below of=x2] {$Y_2$};
\node[main] (y3) [below of=x3] {$Y_3$};
\draw[->] (x1) -- (y1);
\draw[->] (x1) -- (x2);
\draw[->] (x1) to [out=30,in=150] (x3);
\draw[->] (y1) -- (x2);
\draw[->] (y1) -- (y2);
\draw[->] (y1) -- (x3);
\draw[->] (y1) to [out=330,in=210] (y3);
\draw[->] (x2) -- (y2);
\draw[->] (x2) -- (x3);
\draw[->] (y2) -- (x3);
\draw[->] (y2) -- (y3);
\draw[->] (x3) -- (y3);
\end{tikzpicture}
}
\\

\subcaptionbox{\centering Time lag dependence $Y_t \independent (X_1,\dots,X_{t-2}) \mid (X_{t-1}, Y_1, \dots, Y_{t-1})$\label{subfig:DAG3}}[0.47\linewidth]{
\begin{tikzpicture}[node distance={15mm}, main/.style = {draw, circle}] 
\node[main] (x1) {$X_1$}; 
\node[main] (x2) [right of=x1] {$X_2$};
\node[main] (x3) [right of=x2] {$X_3$};
\node[main] (y1) [below of=x1] {$Y_1$};
\node[main] (y2) [below of=x2] {$Y_2$};
\node[main] (y3) [below of=x3] {$Y_3$};
\draw[->] (x1) -- (x2);
\draw[->] (x1) -- (y2);
\draw[->] (x1) to [out=30,in=150] (x3);
\draw[->] (y1) -- (x2);
\draw[->] (y1) -- (y2);
\draw[->] (y1) -- (x3);
\draw[->] (y1) to [out=330,in=210] (y3);
\draw[->] (x2) -- (x3);
\draw[->] (x2) -- (y3);
\draw[->] (y2) -- (x3);
\draw[->] (y2) -- (y3);
\end{tikzpicture}
}
&

\subcaptionbox{\centering Exogenous covariates $X_t \independent (Y_1,\dots,Y_{t-1}) \mid (X_1,\dots,X_{t-1})$\label{subfig:DAG4}}[0.53\linewidth]{
\begin{tikzpicture}[node distance={15mm}, main/.style = {draw, circle}] 
\node[main] (x1) {$X_1$}; 
\node[main] (x2) [right of=x1] {$X_2$};
\node[main] (x3) [right of=x2] {$X_3$};
\node[main] (y1) [below of=x1] {$Y_1$};
\node[main] (y2) [below of=x2] {$Y_2$};
\node[main] (y3) [below of=x3] {$Y_3$};
\draw[->] (x1) -- (y1);
\draw[->] (x1) -- (x2);
\draw[->] (x1) -- (y2);
\draw[->] (x1) to [out=30,in=150] (x3);
\draw[->] (x1) -- (y3);
\draw[->] (y1) -- (y2);
\draw[->] (y1) to [out=330,in=210] (y3);
\draw[->] (x2) -- (y2);
\draw[->] (x2) -- (x3);
\draw[->] (x2) -- (y3);
\draw[->] (y2) -- (y3);
\draw[->] (x3) -- (y3);
\end{tikzpicture}
}
\\

\subcaptionbox{\centering Markov covariates $X_t \independent (X_1, \dots, X_{t-2}) \mid (X_{t-1},Y_1,\dots,Y_t)$\label{subfig:DAG5}}[0.47\linewidth]{
\begin{tikzpicture}[node distance={15mm}, main/.style = {draw, circle}] 
\node[main] (x1) {$X_1$}; 
\node[main] (x2) [right of=x1] {$X_2$};
\node[main] (x3) [right of=x2] {$X_3$};
\node[main] (y1) [below of=x1] {$Y_1$};
\node[main] (y2) [below of=x2] {$Y_2$};
\node[main] (y3) [below of=x3] {$Y_3$};
\draw[->] (x1) -- (y1);
\draw[->] (x1) -- (x2);
\draw[->] (x1) -- (y2);
\draw[->] (x1) -- (y3);
\draw[->] (y1) -- (x2);
\draw[->] (y1) -- (y2);
\draw[->] (y1) -- (x3);
\draw[->] (y1) to [out=330,in=210] (y3);
\draw[->] (x2) -- (y2);
\draw[->] (x2) -- (x3);
\draw[->] (x2) -- (y3);
\draw[->] (y2) -- (x3);
\draw[->] (y2) -- (y3);
\draw[->] (x3) -- (y3);
\end{tikzpicture}
}
&

\subcaptionbox{\centering Markov outcome $Y_t \independent (Y_1, \dots, Y_{t-2}) \mid (Y_{t-1}, X_1,\dots,X_t)$\label{subfig:DAG6}}[0.53\linewidth]{
\begin{tikzpicture}[node distance={15mm}, main/.style = {draw, circle}] 
\node[main] (x1) {$X_1$}; 
\node[main] (x2) [right of=x1] {$X_2$};
\node[main] (x3) [right of=x2] {$X_3$};
\node[main] (y1) [below of=x1] {$Y_1$};
\node[main] (y2) [below of=x2] {$Y_2$};
\node[main] (y3) [below of=x3] {$Y_3$};
\draw[->] (x1) -- (y1);
\draw[->] (x1) -- (x2);
\draw[->] (x1) -- (y2);
\draw[->] (x1) -- (y3);
\draw[->] (x1) to [out=30,in=150] (x3);
\draw[->] (y1) -- (x2);
\draw[->] (y1) -- (y2);
\draw[->] (y1) -- (x3);
\draw[->] (x2) -- (y2);
\draw[->] (x2) -- (x3);
\draw[->] (x2) -- (y3);
\draw[->] (y2) -- (x3);
\draw[->] (y2) -- (y3);
\draw[->] (x3) -- (y3);
\end{tikzpicture}
}
\\

\subcaptionbox{\centering Time dependence, exogenous covariates, Markov covariates and outcome\label{subfig:DAG7}}[0.47\linewidth]{
\begin{tikzpicture}[node distance={15mm}, main/.style = {draw, circle}] 
\node[main] (x1) {$X_1$}; 
\node[main] (x2) [right of=x1] {$X_2$};
\node[main] (x3) [right of=x2] {$X_3$};
\node[main] (y1) [below of=x1] {$Y_1$};
\node[main] (y2) [below of=x2] {$Y_2$};
\node[main] (y3) [below of=x3] {$Y_3$};
\draw[->] (x1) -- (y1);
\draw[->] (x1) -- (x2);
\draw[->] (y1) -- (y2);
\draw[->] (x2) -- (y2);
\draw[->] (x2) -- (x3);
\draw[->] (y2) -- (y3);
\draw[->] (x3) -- (y3);
\end{tikzpicture}
}
&

\subcaptionbox{\centering CEG representation of the DAG in (g) with all binary variables\label{subfig:DAG8}}[0.53\linewidth]{
\includegraphics[scale=0.28]{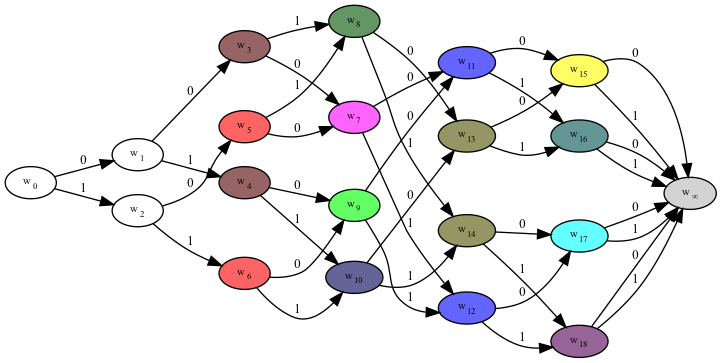} 
}

\end{tabular}
\caption{Possible Markov assumptions and their DAGs for covariates and responses observed at three time points.  Final graph shows the CEG with all binary variables associated to the DAG in (g).}\label{fig:DAGs}
\end{figure}

Although the specific Markov assumptions should be chosen based on the application, the longitudinal nature of the data supports some assumptions that might be relevant to many different applications.  Some possibilities have been summarised in Figure~\ref{fig:DAGs}.

A common assumption in regression models, for example the marginal model, is that the outcome at time $t$ only depends on the time dependent covariates through those at time $t$ (Figure~\ref{subfig:DAG2}).  Alternatively, if $X_t$ and $Y_t$ are measured simultaneously, the effect of the covariates might be lagged so that $Y_t$ only depends on $X_{t-1}$ (Figure~\ref{subfig:DAG3}).  Assumptions can also be made about how future covariates depend on past outcomes.  Most simply, we might assume that they are conditionally independent or exogenous \citep{hernan2001marginal} (Figure~\ref{subfig:DAG4}).  Further possible assumptions are that the outcomes or the covariates follow Markov processes (Figures \ref{subfig:DAG5} and \ref{subfig:DAG6}).

Combining a number of these assumptions can greatly simplify the staged tree.  For example, Figure~\ref{subfig:DAG7} shows a DAG for three time points representing the assumptions of time dependence, exogenous covariates and Markov covariates and outcomes, alongside the equivalent CEG representation of these assumptions with binary variables in Figure~\ref{subfig:DAG8}.  This CEG, for example, has combined the 32 situations associated to the final outcome $Y_3$ into only 4 stages.

\begin{example}
A simple assumption we could make in the depression example is that the outcomes follow a Markov process.  This corresponds to placing certain vertices in the final level of the tree in the same stage.
Therefore the sample size can be shared across such vertices.  For example, returning to Figure~\ref{fig:DepET}, the situation $s_{30}$ has a sample size of only 4.  But with the Markov assumption, $s_{30}$ is in the same stage as $s_{28}$ and their combined sample size is 28.


With this Markov assumption, the selected CEG (shown in Figure~\ref{fig:DepCEGMarkov}) has been simplified somewhat with each of the four covariate combinations being separated.  Furthermore, for patients using the new drug or patients using the standard treatment with mild symptoms, there is a context specific independence between the outcomes in weeks 2 and 4.

\begin{figure}[h]
    \centering
    \includegraphics[scale=0.405]{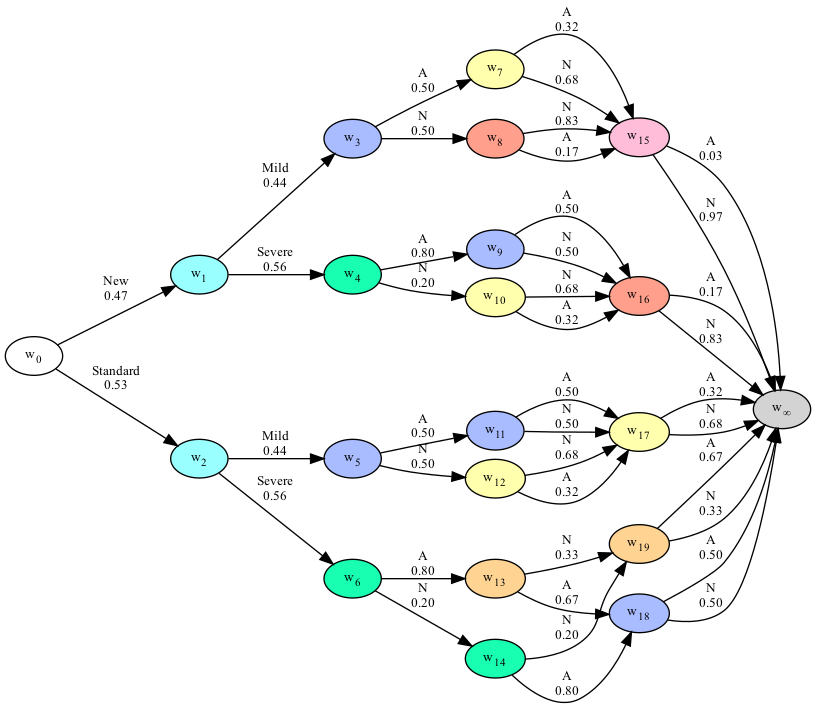}
    \caption{CEG for depression example with Markov process outcomes assumption.}
    \label{fig:DepCEGMarkov}
\end{figure}

\end{example}

\subsection{Marginal longitudinal STs}\label{sec:Margins}

In the previous section, Markov assumptions were used to improve inferences in cases where sample sizes are too small for the given event tree.  However, when the event tree is large, the graphical representation of the ST or CEG can be hard to interpret making inferences challenging to read.  Instead it may be useful to consider certain marginal distributions which generate event trees of smaller size making the resulting CEG easier to interpret.  It also reduces the number of parameters in the model, improving estimation is cases of small sample sizes.

The choice of marginal distributions depends on the application and the aims of the analysis.  For example, if one is interested in the marginal distributions at each time point we might fit different STs for each $(Z,X_t,Y_t)$, $t=1,\dots,T$.  Alternatively, if one is interested in the long term effects of the covariates on the final outcome, we might instead consider the margins $(Z,X_t,Y_t,Y_T)$, $t=1,\dots,T-1$.

A limitation to this approach is that it does not directly use the longitudinal nature of the data and therefore misses out on its potential benefits.  For example, the marginal regression model is able to utilise the whole data when estimating the marginal probabilities by specifying a covariance structure between the repeated observations.  While doing this for STs is more difficult - unlike in regression models, a ST explicitly models the covariates and so a covariance structure is required for all timed covariates and outcomes - there are some alternative ways to incorporate the longitudinal nature of the data into STs.

Often in longitudinal data the covariates and outcomes at different times share the same sample space and so the event trees for each of the margins share the same topology.  One can therefore conduct further analysis by comparing the stagings of the selected STs for the different margins.  This can give an idea of how the relationship between the outcome and covariates changes over time.

When the outcome variable is numeric, another way to incorporate the longitudinal nature of the data is to redefine the outcome variables as the difference between the outcome at different time points.  This was the approach taken by, for example, \cite{Scutari2017Bayesian} when using BNs to analyse longitudinal data.  This focuses the analysis on how the outcome changes over time, rather than on its numeric value.  An example of this will be given in Section \ref{sec:Caries}, but first we return to the depression example.

\begin{example}

    We consider the marginal distribution of the covariates with each of the three timed outcomes individually to uncover the marginal effect of the two treatments at different times.  Note that now the smallest sample size for a situation is 70.  The selected CEGs for the three margins are in Figure~\ref{fig:DepCEGMargin}.

    \begin{figure}
        \centering
        \begin{tabular}[c]{c} 
        \subcaptionbox{Week 1}
        {\includegraphics[scale=0.5]{CEG12.png}} \\ 
        \subcaptionbox{Week 2}
        {\includegraphics[scale=0.5]{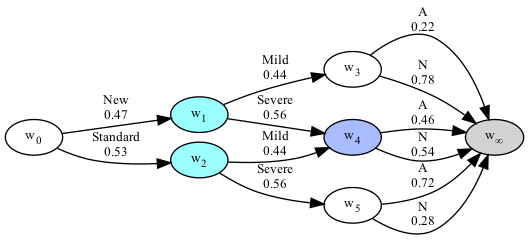}} \\ 
        \subcaptionbox{Week 4}
        {\includegraphics[scale=0.5]{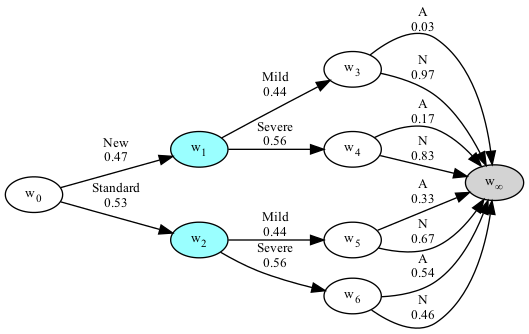}}
        \end{tabular}
        \caption{Marginal CEGs for depression example.}
        \label{fig:DepCEGMargin}
    \end{figure}

    In week 1, the new drug and standard treatment are placed in the same position - this means that in the model there is no difference between the treatments at this time.  In week 2 the treatments are no longer in the same position.  However, the new drug with severe symptoms is in the same stage at the standard treatment with mild symptoms, both with a 0.54 probability of normal symptoms.  The difference between the two treatments is even greater in week 4 when all situations before the outcome are in different stages.  The new treatment leads to higher probability of normal symptoms in week 4 no matter the diagnosis severity. 

\end{example}

\section{Comparison to regression methods and benefits of STs}\label{sec:Comparison}

In this section we compare the proposed ST methods to regression based methods and discuss some of the benefits of using STs for longitudinal data.  We start by continuing the depression example, comparing the estimated probabilities obtained by the ST methods in the previous sections to a marginal regression model.  We then give a more general discussion on the differences between ST models and regression models, focusing on the different assumptions they make and the conclusions at which they are able to arrive.  But first we give a brief introduction to three standard regression based approaches for  modelling longitudinal data with linear models - the marginal model, transition model and random effects model \cite[see][Chapters~7~-~10]{diggle2002analysis}.  The primary difference between these three models is in how they characterise the dependence between the outcomes at different time points.

The marginal model models the regression and longitudinal dependence separately.  A common regression model is assumed over each time point with outcome $Y_t$ and covariates $Z,X_t$.  Although the model is assumed to be the same at each time point, differences between the time points can be expressed through the inclusion of time $t$ as a covariate, as well as interaction terms between $t$ and the other covariates.  Defining the likelihood function also requires a model for the dependence between the outcomes $Y_1,\dots,Y_T$.  With discrete outcomes, this is a complex task which involves many nuisance parameters.  Instead, a generalised estimating equations approach only requires a model for the pairwise correlations.  Common choices for this correlation model are an exchangeable structure where $\mathrm{Corr}(Y_s,Y_t)$ is identical (but unknown) for all $s,t$ or an autoregressive structure.

The transition model models both the effect of the covariates on the outcome and the time dependence within the same logistic regression.  This is achieved by including functions of the previous outcomes $Y_1,\dots,Y_{t-1}$ as predictors in the logistic regression for $Y_t$.  This allows for a more nuanced description of the time dependence than in the marginal model, however additional care must be taken in the interpretation of the regression coefficients because their value depends the chosen functions of previous outcomes.

The random effects model takes a different approach by allowing some regression coefficients $\beta_i$ to vary across individuals $i$.  It is then assumed that any dependence between the longitudinal outcomes is captured by $\beta_i$ - that is, $Y_1,\dots,Y_T$ are conditionally independent given $\beta_i$.  Additionally, it is required to specify a distribution for the coefficients $\beta_i$ - a common choice is a zero mean Gaussian distribution.  This approach is useful because it is allows inference on the individual level rather than just the population level.

\subsection{Depression example}\label{subsec:DepExample}

The depression data set was analysed using a marginal regression model in \cite{agresti2012categorical}, Chapter 9.  The three time points are labelled $t=0,1,2$.  The outcome variables are $Y_0,Y_1,Y_2$ with $Y_t=1$ denoting normal symptoms and $Y_t=0$ abnormal symptoms.  The treatment allocation is denoted by $d = 0$ for the standard treatment and $d=1$ for the new drug, and the diagnosis severity by $s=0$ for mild and $s=1$ for severe.  The logistic regression model used by \cite{agresti2012categorical} was
\begin{equation*}
    \mathrm{logit}\left( P(Y_t=1 \mid s,d) \right) = \alpha + \beta_1 s + \beta_2 d + \beta_3 t + \beta_4 d t
\end{equation*}
Additionally, an exchangeable correlation structure is used so that $\mathrm{Corr}(Y_s,Y_t) = \rho$ for all pairs $s,t$.  Parameter estimates are obtained using the generalised estimating equation method.

The fitted correlation between different times was very weak with estimated $\hat{\rho} = -0.003$.  This is somewhat reflected in the CEG in Figure~\ref{fig:DepCEG}, however the CEG also highlights that this independence is context specific - it only occurs for certain combinations of the covariates and previous outcomes.  For example, $Y_1$ is independent of $Y_0$ when $d=1$ and $s=1$.  However, they are not independent for any of the other combinations of covariates.

The general conclusions from the estimated regression coefficients also match the conclusions from the ST models.  The estimated time effects are $\hat{\beta}_3=0.482$ for the standard treatment and $\hat{\beta}_3+\hat{\beta}_4=1.500$ for the new drug.  The Wald test for $\beta_4=0$ has p-value $< 0.0001$ demonstrating strong evidence for a faster improvement in symptoms when using the new drug.  Additionally, the estimated effect of the new drug at time $t=0$ is $\hat{\beta}_2=-0.060$ - a very small effect demonstrating an insignificant difference between the treatments in the first week.  Both of these conclusions match those of all three ST methods.

Estimated marginal probabilities for the outcome at each time point conditional on the covariates can be found in Table \ref{tab:probs} for each of the three ST methods and regression model.  While these probabilities are generally very similar, there are some notable differences.  Between the ST methods, differences in estimated probabilities occur due to different stagings in the selected models.  For example, for $d=0,s=0$ the estimate of the probability of $Y_2=1$ is 0.62 under the full ST, but only 0.54 under the marginal ST.  This is because the marginal ST includes this situation in the same stage as $d=1,s=1$.

\begin{table}[h]
\begin{tabular}{ll|lll}
Covariates                                  & Model        & Week 1 & Week 2 & Week 4 \\ \hline
\multirow{4}{*}{New drug, severe}           & Full ST      & 0.24   & 0.51   & 0.83   \\
                                            & Markov ST    & 0.20   & 0.54   & 0.83   \\
                                            & Marginal STs & 0.20   & 0.54   & 0.83   \\
                                            & Regression   & 0.20   & 0.52   & 0.83   \\ \hline
\multirow{4}{*}{New drug, mild}             & Full ST      & 0.51   & 0.80   & 0.95   \\
                                            & Markov ST    & 0.50   & 0.76   & 0.97   \\
                                            & Marginal STs & 0.52   & 0.78   & 0.97   \\
                                            & Regression   & 0.48   & 0.80   & 0.95   \\ \hline
\multirow{4}{*}{Standard treatment, severe} & Full ST      & 0.24   & 0.24   & 0.46   \\
                                            & Markov ST    & 0.20   & 0.30   & 0.45   \\
                                            & Marginal STs & 0.20   & 0.28   & 0.46   \\
                                            & Regression   & 0.21   & 0.30   & 0.41   \\ \hline
\multirow{4}{*}{Standard treatment, mild}   & Full ST      & 0.51   & 0.62   & 0.65   \\
                                            & Markov ST    & 0.5    & 0.59   & 0.68   \\
                                            & Marginal STs & 0.52   & 0.54   & 0.67   \\
                                            & Regression   & 0.49   & 0.61   & 0.72  
\end{tabular}
\caption{Estimated marginal probabilities of normal symptoms for each time point conditional on covariate values}\label{tab:probs}
\end{table}

On the other hand, the regression model can obtain different probability estimates due to the linear assumptions in the model.  In particular, it assumes that the probabilities change linearly over time with respect to the logit link function.  This results in the estimated probabilities for $d=0,s=0$ of 0.49, 0.61 and 0.72 for weeks 1,~2~and~4 respectively.  In comparison, the full ST model estimates these probabilities as 0.51, 0.62 and 0.65.  Notice the large difference in estimated probabilities at week 4.  Free from the linearity assumption, the ST method is able to estimate a lower probability demonstrating diminishing returns from the standard treatment over time.

The marginal STs and regression model are only able to estimate marginal probabilities of the outcome at each time.  For the regression model, this is because the marginal probabilities and pairwise correlations do not specify the joint distribution of the outcomes.  On the other hand, the full ST and Markov ST methods are able to estimate the full joint probabilities.  This enables slightly more nuanced observations.  For example, for $d=1,s=1$, all methods estimate the probability of normal symptoms at week 4 to be $0.83$.  However, with the full ST model we can see that this probability is estimated to be 0.86 if the patient had abnormal symptoms at week 1 but only 0.73 if they had normal symptoms at week 1.  Such context specific information could be used to reassure a patient that was worried after continuing to experience abnormal symptoms after one week of treatment. 

\subsection{Model assumptions}

The full ST described in Section \ref{sec:FullST} makes no assumptions about the data beyond the sample space described by the event tree.  This lack of assumptions can be a benefit because it makes it applicable to any appropriate data set and the conclusions are not dependent on any assumptions.  However, as has been mentioned previously, this lack of assumptions requires a large sample size relative to the size of the event tree in order for probability estimation and model selection to be performed reliably. This is because the full joint probability distribution needs to be estimated and, as the event tree gets larger, this joint distribution involves many parameters.  In situations where there is not sufficient data, one can reach conclusions more supported by data by making additional assumptions - although the conclusions are now dependent on these assumptions being true.

Regression based models can make a number of assumptions depending on the approach taken.  Common to all regression methods is that the outcome probabilities are linear functions of the covariates (with respect to a chosen link function).  While this assumption can be relaxed slightly by including interaction terms between the covariates, it still results in the effect of the regression coefficients being additive.

Additionally, regression based methods make assumptions about the dependence between the outcome at different time points.  The marginal model does this separately from the regression by assuming some correlation structure between the outcomes at different times.  The transition model includes the previous outcomes in the regression resulting in an additive effect.  The random effects model assumes independence of the outcomes conditional on some individual level parameter.  Depending on the application, these assumptions may be more or less appropriate.

In Section \ref{sec:Markov} we suggested the use of ST models with Markov assumptions given by a DAG which specify conditional independences between the variables.  However, beyond these conditional independences, no additional restrictions are placed on the model.  The choice of Markov assumptions is left to the practitioner, allowing the assumptions to be chosen to best fit the specific application.  Additionally, due to the DAG respresentation, these assumptions are easy to interpret and can be verified either using expert judgement or testing the conditional independences using data.

Like the full ST model, the marginal ST models of Section \ref{sec:Margins} make no assumptions about the data.  However, they only model the marginal distributions rather than the full joint distribution so the conclusions one can draw are weaker.  However, when combined with certain (conditional) independence assumptions, one could construct the full joint probability distribution.  For example, the marginal STs for each individual time point $(Z,X_t,Y_t)$, $t=1,\dots,T$ combined with the assumption that the time points are independent given $Z$ results in the full joint probability distribution.

\subsection{Model interpretation}

The primary interest of regression based models is the effect of the covariates on the outcome.  As such, they only model the conditional distribution of the outcome given the covariates.  One is able to estimate these conditional distributions via the estimates for the regression coefficients.  One can also use these regression coefficients to infer the effect of the covariates on the outcome.  In the marginal model, interpretation of the regression coefficients is simple and analogous to a standard logistic regression.  However, in the transitions model and random effects model this interpretation is more complicated since the value of the regression coefficients depends on other modelling choices \citep{zeger1992overview}.

By including interaction terms in the regression between the covariates and time, one is able to determine how the effect of a covariate changes over time.  However, this change is restricted to be linear (and therefore monotonic) over time.  By testing for coefficients being equal to zero (either using p-values or more sophisticated model selection methods), one can also find which covariates have no effect on the outcome, or which effects do not change over time.

One can also use regression methods to infer the dependence between the different timed outcomes.  How this is done depends on the specific method - the marginal model estimates a common covariance function between all outcomes while the transition model interprets the dependence through additional regression coefficients.

The conclusions made by the proposed ST methods are twofold.  First there is the model selection step in which the ST with highest posterior probability that best describes the data is searched for.  While there are often many models with high posterior probability \citep{strong2022bayesian}, the selected model can still provide a useful interpretation of the data.  For example, by identifying situations that are in the same stage, one can find different combinations of the covariates which have the same effect on the outcome in terms of the probability distribution.  

Second, given the selected ST, transition probabilities are estimated.  In this regard ST methods go one step further than regression methods because they explicitly model the covariates.  This means that conclusions are not limited to the outcome variable but can also be made about the covariates - the dependence between the covariates and how they change over time.  This can be particularly useful in a longitudinal study because if there is a covariate which has a strong effect on the outcome, it would also be important to understand how the value of this covariate changes over time.

\subsection{Other benefits of STs}

Aside from the differences discussed above, ST models also enjoy some additional benefits in the modelling of longitudinal data.

A key benefit of ST models, not limited to longitudinal data, is the graphical representation.  The CEG provides a visual aid for interpreting the conclusions of the model selection which can be understood without expertise in how the model works.  The event tree also provides a useful visualisation tool when the edges of the event tree are labelled with the number of observations in the data set.  Using this, one can easily identify situations in the tree that have few or zero observations.  This is important because any conclusions or probability estimation made about such small sample size situations will be based primarily on the modelling assumptions.  In a ST modelling process, such limitations of the data set would become obvious when drawing the event tree and can be made evident in the selected CEG by using the new representation suggested at the end of Section \ref{sec:STs}.  However, in standard regression modelling this would be harder to identify without careful exploratory analysis.  In particular, the linear assumptions of regression models mean that strong inferences can be made about situations with few observations.

STs can be updated with only a partial observation of the variables \citep{freeman2011bayesian}.  This is useful because data points which only observe the variables up to a certain time point - either due to dropout or because the individual has not completed all time points - can still be used to update probabilities.

In this paper we have distinguished the outcome variable from the covariates.  Such a distinction is useful when the focus of the analysis is on the effect of the covariates on the outcome, as is the case in regression models.  However, this distinction is not necessary for ST models allowing for a more general modelling approach.  This can be useful when one is instead interested in the distribution of all observed variables.  Each of the three ST methods in this paper remain valid in this case.  

In fact STs are able to generalise further.  So far in this paper we have focused on stratified event trees in which each layer of the tree can be associated to a variable.  However, a key benefit of ST models is that they do not require a variable based representation and can model asymmetric or \textit{non-stratified} processes.  This is useful in data sets which contain structural zeros - when some sequence of events occurs with probability zero - or when the sample space of an event depends on the previous events.  This might occur in a longitudinal data set when what is measured at time $t$ depends on the observed values at the previous times.  


\section{Dentistry data example}\label{sec:Caries}

We now demonstrate the proposed ST methods on a longitudinal data set examining the prevalence of caries (also called tooth decay or cavities) in children.  Initially a cross-sectional study was conducted and analysed by \cite{ugolini2018trends} and, based on the findings of this study, a further longitudinal study was carried out.  An initial version of this longitudinal study with three time points was analysed by \cite{ugolini2023probabilistic} using an undirected graphical model.  Here we have access to an extended version of this data set with four time points with time-dependent variables measured at ages 3, 4, 5 and 7.  We denote these four time points by $t=1,2,3,4$.

In \cite{ugolini2023probabilistic}, the selected undirected graphical model found a direct relationship between a child's oral hygiene and the prevalence of caries.  There was also an edge between the length of breastfeeding time and oral hygiene, however the prevalence of caries was found to be independent of breastfeeding time conditional on oral hygiene.  In this graphical model, repeated observations were summarised into a single variable and so it was more similar to a cross-sectional or marginal analysis.  We aim to investigate the relationship between these three variables over the whole time series.

The data set contains the following variables:
\begin{itemize}
    \item Breastfeeding time $B$.  An indicator of short breastfeeding time of 0-9 months (including no breastfeeding) or long breastfeeding time of 10+ months.  The cutoff of 9 months was chosen to obtain two balanced classes.
    \item Oral hygiene $H_1,\dots,H_4$.  A clinical assessment of whether the child's oral hygiene was `adequate' or `inadequate' measured at all four time points $t=1,\dots,4$.
    \item Change in caries incidences $Y_{ij}$.  At each age the total number of caries incidences (the sum of cavities and fillings) was recorded, denoted by $I_1,\dots,I_4$.  As the outcome variable, we are interested in whether the number of incidences increases or not between different ages.  These are denoted by the variables $Y_{ij}$, $i,j=1,\dots,4$, $i<j$ where $Y_{ij}$ is `increasing' if $I_j - I_i > 0$ and is `decreasing' if $I_j - I_i \leq 0$ (no change is included in the decreasing category).  At the first recorded time point at age 3, the outcome is instead denoted by either $Y_1$ or $Y_{01}$ and is `caries' if $I_1 > 0$ and `no caries' if $I_1=0$.
\end{itemize}

The data set contains 277 participants of which 237 have full observation of all variables (missing values were due to a seasonal flu and so can be considered missing at random). 
Modelling this data with a full ST model might consider the nine binary variables $(B,H_1,Y_1,H_2,Y_{12},H_3,Y_{23},H_4,Y_{34})$.  The corresponding event tree has 512 paths and with only 237 full observations, the sample size is not sufficient to reliably estimate the full joint probability.  In fact, there are only observations for 37 of the 512 paths and 135 of the observations are on the two paths with all adequate hygiene and no caries.  This indicates that modelling the data with the full longitudinal ST with no additional assumptions is not realistic.

Instead, as suggested in Section \ref{sec:Margins}, we first consider certain marginal distributions of the data.

\subsection{Marginal STs}\label{subsec:DenMargin}

\begin{figure}[h]
    \centering
    \begin{tabular}[c]{c} 
    \subcaptionbox{Age 3}
    {\includegraphics[scale=0.42]{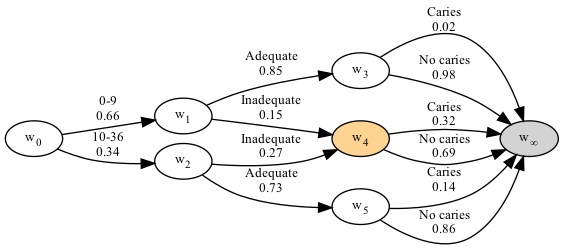}} \\
    \subcaptionbox{Age 4}
    {\includegraphics[scale=0.42]{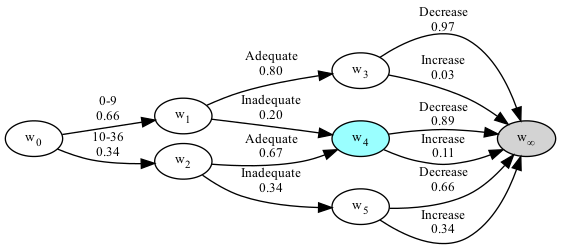}} \\
    \subcaptionbox{Age 5}
    {\includegraphics[scale=0.42]{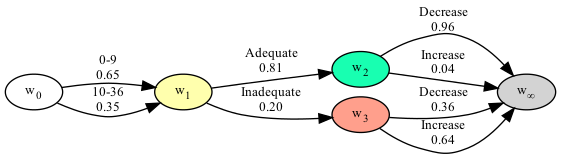}} \\
    \subcaptionbox{Age 7}
    {\includegraphics[scale=0.42]{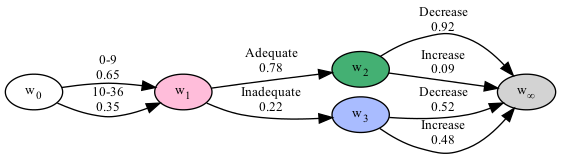}}
    \end{tabular}
    \caption{Marginal CEGs at each time point $(B,H_t,Y_{t-1,t})$.}
    \label{fig:DenMar}
\end{figure}

The first marginal distributions we investigate are for each individual time point $(B,H_t,Y_{t-1,t})$.  The selected CEGs are in Figure~\ref{fig:DenMar}.  At ages 3 and 4, a long breastfeeding time is associated with a higher probability of inadequate hygiene and, in some cases, a higher probability of increasing caries incidences.  However, at ages 5 and 7 breastfeeding time is marginally independent of oral hygiene and incidence change.

\begin{figure}[h]
    \centering
    \begin{tabular}[c]{c} 
        \subcaptionbox{Age 3}
        {\includegraphics[scale=0.35]{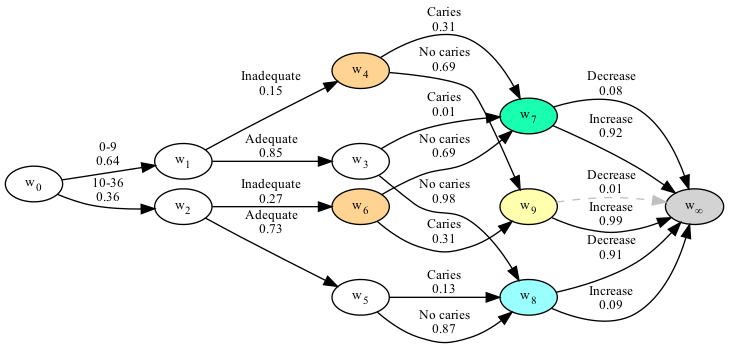}} \\
        \subcaptionbox{Age 4}
        {\includegraphics[scale=0.35]{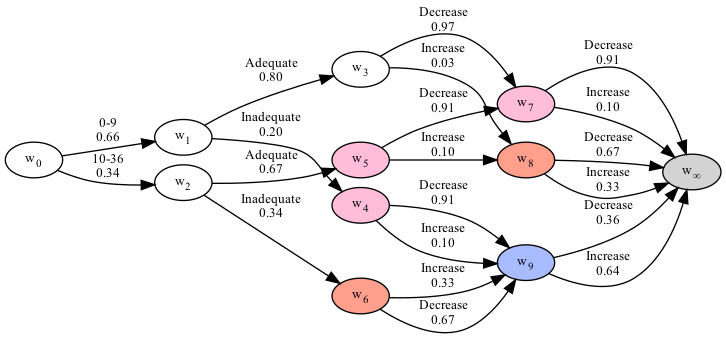}} \\
        \subcaptionbox{Age 5}
        {\includegraphics[scale=0.35]{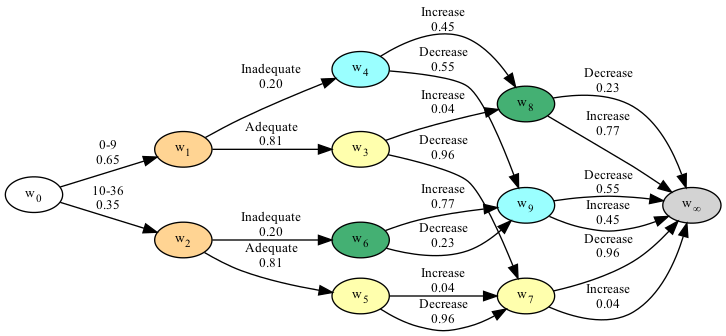}}
    \end{tabular}
    \caption{Marginal CEGs for each time point with change to age 7 $(B,H_t,Y_{t-1,t},Y_{t7})$.}
    \label{fig:DenAge7}
\end{figure}

 For a longer term effect of the covariates, the change in incidences to the final time point can be included at the end of each marginal CEG $(B,H_t,Y_{t-1,t},Y_{t7})$.  The selected CEGs are in Figure~\ref{fig:DenAge7}.  In all selected CEGs, paths with inadequate hygiene lead to situations with higher probability of increasing incidences at age 7, while paths with adequate hygiene generally lead to lower probability.  However, for ages 3 and 5 the path with short breastfeeding time, adequate hygiene and increasing incidences arrives at a situation with a high probability of increasing incidences at age~7.

In the first marginal CEGs of Figure~\ref{fig:DenMar}, we found evidence that inadequate hygiene is a risk factor for caries.  Additionally, there is evidence that at younger ages a long breastfeeding time can increase the probability of inadequate hygiene, indirectly leading to higher risk of caries.  There is also slightly weaker evidence that long breastfeeding time is a direct risk factor for caries at younger ages.  However, in the marginal CEGs of Figure~\ref{fig:DenAge7}, it was found that individuals that have increasing caries incidences despite having neither of these two risk factors (i.e. short breastfeeding time and adequate hygiene) have a high probability of increasing incidences in the future.  This perhaps suggests the presence of a further risk factor not included in this data set - although it should be noted that this is based on relatively few observations.

\begin{figure}[h]
    \centering
    \begin{tabular}[c]{c} 
        \includegraphics[scale=0.36]{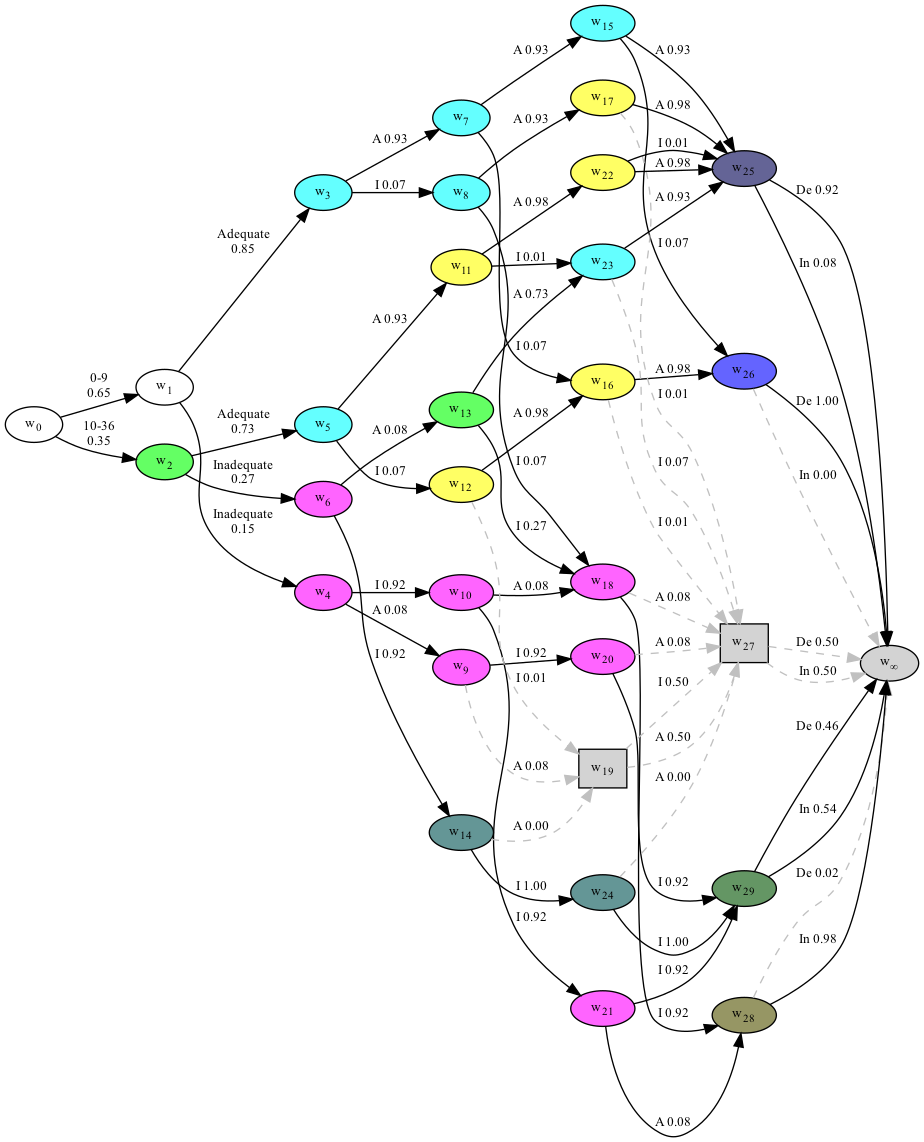}
    \end{tabular}
    \caption{Marginal CEG for incidence change at age 7 with all past covariates $(B,H_1,H_2,H_3,H_4,Y_{34})$.  Some edge labels are abbreviated: Adequate~(A), Inadequate~(I), Increase~(In), Decrease~(De).}
    \label{fig:DenFullCovariates}
\end{figure}

The marginal CEGs of Figure~\ref{fig:DenAge7} further show that inadequate hygiene at younger ages is associated with a higher probability of increasing incidences at older ages.  Within this data set, there are two likely reasons for this.  One is that hygiene at different ages are highly correlated, i.e. inadequate hygiene at a younger age is associated to inadequate hygiene at older ages which is in turn a risk factor for caries at older ages.  The other is that inadequate hygiene at younger ages is a direct risk factor for caries at older ages.  To investigate this further we consider the marginal CEG for the increase in incidences at age 7 with all covariates $(B,H_1,H_2,H_3,H_4,Y_{34})$.  The selected CEG is in Figure~\ref{fig:DenFullCovariates}.  In this CEG, there are two stages with low probability of increasing incidences - $w_{25}$ and $w_{26}$.  All paths leading to these stages have either 0 or 1 occurrence of inadequate hygiene throughout the time series.  The other two stages $w_{28}$ and $w_{29}$ both have higher probability of increasing incidences and all paths leading to these stages have 3 or 4 occurrences of inadequate hygiene.  A similar phenomenon was also observed for age 5 but this CEG is omitted for brevity.  Based on this, we make the hypothesis that the number of occurrences of inadequate hygiene is a risk factor for caries, rather than the exact time that they occur.  Note that this is a context specific property which could not be discovered directly from a BN.

\subsection{ST with Markov assumptions}

We now select a ST for the full data set $(B,H_1,Y_1,H_2,Y_{12},H_3,Y_{23},H_4,Y_{34})$ with additional assumptions (Section \ref{sec:Margins}).  These assumptions are in part informed by the conclusions of the marginal STs in Section \ref{subsec:DenMargin}.  Thus, to specify the assumptions we consider a DAG using redefined hygiene variables that count the number of inadequate occurrences - $H'_t = \sum_{i=1}^{t} H_i$.  We refer to these as the hygiene sums.  Note that any BN using the hygiene sums can still be expressed using a ST model on the original hygiene variables.

Our chosen Markov assumptions are represented in the DAG in Figure~\ref{fig:DenBN2} and can be summarised as:
\begin{itemize}
    \item Ages 4, 5 and 7 are conditionally independent of breastfeeding time given age 3.
    \item Hygiene sums follow a Markov process and are exogenous.
    \item Incidences follow a Markov process and only directly depend on hygiene through the sum at the corresponding time.
\end{itemize}

To further validate these assumptions, we select a BN model from the data using the R package \texttt{bnlearn} \citep{Scutari2010learning}.  We first specified that the direction of edges in the BN should respect the chronology of the data.  Then, to select between different models we used the Akaike Information Criterion and used Tabu search to explore the model space for the highest scoring model.  The selected BN is in Figure~\ref{fig:DenBN1}.  The selected BN closely matches our chosen BN, lending strength to the assumptions that were made.

\begin{figure}
    \centering
    \begin{tabular}[c]{c} 
        \subcaptionbox{DAG representing chosen assumptions in ST model\label{fig:DenBN2}}
        {\begin{tikzpicture}[node distance={15mm}, main/.style = {draw, circle}] 
            \node[main] (B) {$B$}; 
            \node[main] (H1) [above right of=B]{$H'_1$}; 
            \node[main] (H2) [right of=H1] {$H'_2$};
            \node[main] (H3) [right of=H2] {$H'_3$};
            \node[main] (H4) [right of=H3] {$H'_4$};
            \node[main] (Y1) [below right of=B] {$Y_1$};
            \node[main] (Y2) [right of=Y1] {$Y_{12}$};
            \node[main] (Y3) [right of=Y2] {$Y_{23}$};
            \node[main] (Y4) [right of=Y3] {$Y_{34}$};
            \draw[->] (B) -- (H1);
            \draw[->] (B) -- (Y1);
            \draw[->] (H1) -- (Y1);
            \draw[->] (H1) -- (H2);
            \draw[->] (H2) -- (Y2);
            \draw[->] (Y1) -- (Y2);
            \draw[->] (H2) -- (H3);
            \draw[->] (H3) -- (Y3);
            \draw[->] (Y2) -- (Y3);
            \draw[->] (H3) -- (H4);
            \draw[->] (H4) -- (Y4);
            \draw[->] (Y3) -- (Y4);      
        \end{tikzpicture}}
        \\
        \subcaptionbox{DAG selected by data\label{fig:DenBN1}}
        {\begin{tikzpicture}[node distance={15mm}, main/.style = {draw, circle}] 
            \node[main] (B) {$B$}; 
            \node[main] (H1) [above right of=B]{$H'_1$}; 
            \node[main] (H2) [right of=H1] {$H'_2$};
            \node[main] (H3) [right of=H2] {$H'_3$};
            \node[main] (H4) [right of=H3] {$H'_4$};
            \node[main] (Y1) [below right of=B] {$Y_1$};
            \node[main] (Y2) [right of=Y1] {$Y_{12}$};
            \node[main] (Y3) [right of=Y2] {$Y_{23}$};
            \node[main] (Y4) [right of=Y3] {$Y_{34}$};
            \draw[->] (B) -- (H1);
            \draw[->] (B) -- (Y1);
            \draw[->] (H1) -- (Y1);
            \draw[->] (H1) -- (H2);
            \draw[->] (H1) -- (Y2);
            \draw[->] (H2) -- (H3);
            \draw[->] (H3) -- (Y3);
            \draw[->] (Y2) -- (Y3);
            \draw[->] (H3) -- (H4);
            \draw[->] (H4) -- (Y4);
            \draw[->] (Y3) -- (Y4);  
        \end{tikzpicture}}
        \end{tabular}
\caption{DAGs for dentistry data with redefined hygiene variables $H'_t = \sum_{i=1}^{t} H_i$.}
    \label{fig:DenBNs}
\end{figure}

\begin{figure}
    \centering
    \begin{tabular}[c]{c} 
        \includegraphics[scale=0.487,angle=90]{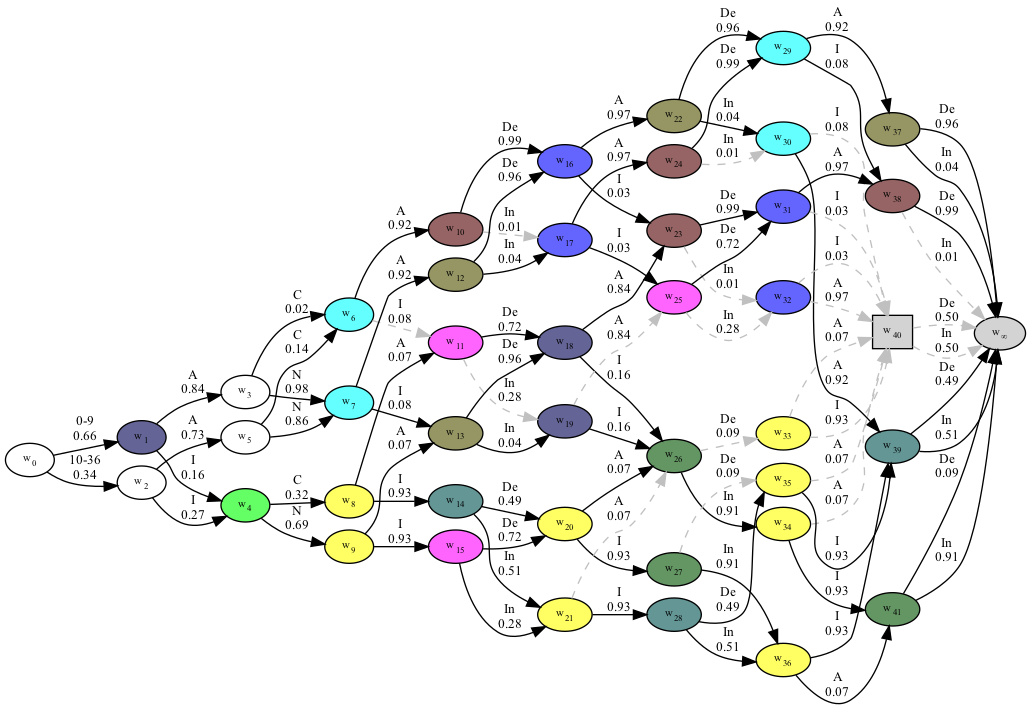}
    \end{tabular}
    \caption{Full CEG for the dentisty data with the Markov assumptions of Figure~\ref{fig:DenBN2}.  Edge labels are abbreviated - Adequate~(A), Inadequate~(I), Caries~(C), No caries~(N), Increase~(In), Decrease~(De).}
    \label{fig:DenFullMarkov}
\end{figure}

The chosen assumptions can now be specified in a ST model and additional model selection conducted.  The selected CEG is in Figure~\ref{fig:DenFullMarkov}.  Despite there being no observations on the majority of paths through the tree, with the benefit of the Markov assumptions there are only 4 situations that have zero sample size which are all associated to the outcome at age 7 (these have been combined into a single zero sample size vertex labelled $w_{40}$).  The CEG can be used to calculate estimated joint probabilities or to track a child through the CEG and provide estimated probabilities for future caries incidences.  

The addition of the Markov assumptions has allowed probability estimation for situations that have few or zero observations.  For example, in the selected CEG $w_{32}$ has no observations.  However, because it is in the same stage as $w_{31}$, we are able to estimate the probability of adequate hygiene at 0.97.

\section{Discussion}

In this paper we have discussed various approaches for using STs to model discrete longitudinal data.  Each of these take advantage of existing methodology for ST model selection and probability estimation and so can be easily implemented - namely by using the Python package \texttt{cegpy} or the R package \texttt{stagedtrees}.  

A main advantage of the proposed ST approaches are the simple assumptions that they make.  Both the full longitudinal ST and marginal longitudinal STs do not make any assumptions about the data.  The longitudinal ST with Markov assumptions only makes assumptions about the conditional independence of the variables (although more complicated assumptions such as context specific independencies are also possible).  Such assumptions are easy to interpret and have a visual representation through a DAG.  Additionally, the assumptions can be verified through standard methods for BN model selection.  This is in contrast to regression methods whose assumptions are often less explicit and harder to interpret.

Another contribution of this paper is the new representation of a CEG for zero sample size situations.  Creating a separate stage for zero sample size situations allows for a more compact and readable representation of the CEG while also providing additional information about the data set and estimated probabilities.  Specifically, it highlights any situations with zero sample size so that one knows that probability estimation for that situation is based purely on the prior distribution and not on data.  This could also  be used to inform new data collection.

Future work in this area could focus on the development of methodology, for example structural learning algorithms, that are tailored to longitudinal data taking into account the dependence between different time points.  One possibility for this is to define an event tree for the whole data set but with a different branch of the tree for each time point.  Another is to assume that the marginal ST remains the same at each time point so that one can use the whole data to select this marginal ST.  An issue with these approaches, similar to the marginal regression model, is that observations are not independent due to the dependence between different time points for the same individual.  Defining the likelihood function therefore requires specification of this dependence between different time points.  This is even more complicated for ST models because the covariates are explicitly modelled - hence the dependence must be specified for all timed covariates as well as the outcome.  However, marginal regression models simplify this process significantly through the use of generalised estimating equations which don't require the full likelihood.  An interesting line of future research would be in the use of the generalised estimating equations for ST models.


\section*{Acknowledgements}
Thank you to Gareth Walley and Aditi Shenvi for their valuable help in using and updating the \texttt{cegpy} package.

The research by JSC and ER was supported in part by the MIUR Excellence Department Project awarded to Dipartimento di Matematica, Università di Genova, CUP D33C23001110001 and the 100021-BIPE 2020 grant.

\section*{Conflict of Interest Statement}
On behalf of all authors, the corresponding author states that there is no conflict of interest.

\bibliography{sn-bibliography}

\end{document}